\documentclass[conference]{IEEEtran}
\IEEEoverridecommandlockouts
\usepackage{cite}
\usepackage{amsmath,amssymb,amsfonts}
\usepackage{hyperref}
\usepackage{algorithm}
\usepackage{dsfont}
\usepackage{algpseudocode}
\usepackage{graphicx}
\usepackage{textcomp}
\usepackage{braket}
\usepackage{xcolor}
\def\BibTeX{{\rm B\kern-.05em{\sc i\kern-.025em b}\kern-.08em
    T\kern-.1667em\lower.7ex\hbox{E}\kern-.125emX}}

\begin{document}
% \makeatletter
% \def\endthebibliography{%
%   \def\@noitemerr{\@latex@warning{Empty `thebibliography' environment}}%
%   \endlist
% }
% \makeatother

\title{QuYBE - An Algebraic Compiler for Quantum Circuit Compression\\
{\footnotesize %\textsuperscript{*}Note: Sub-titles are not captured in Xplore and
%should not be used
}
}

\author{\IEEEauthorblockN{Sahil Gulania$^{\dagger}$}
\IEEEauthorblockA{\textit{Mathematics and Computer Science Division} \\
\textit{Argonne National Laboratory}\\
Lemont, Illinois, 60439, United States \\
sgulania@anl.gov}
\and
    \IEEEauthorblockN{Zichang He$^{\dagger}$}
\IEEEauthorblockA{\textit{Department of Electrical and Computer Engineering} \\
\textit{University of California, Santa Barbara}\\
Santa Barbara, California, 93106, United States \\
zichanghe@ucsb.edu}
\and
\IEEEauthorblockN{Bo Peng$^{\dagger}$}
\IEEEauthorblockA{\textit{Physical and Computational Sciences Directorate} \\
\textit{Pacific Northwest National Laboratory}\\
Richland, Washington, 99352, United States \\
peng398@pnnl.gov}
\and
\IEEEauthorblockN{Niranjan Govind}
\IEEEauthorblockA{\textit{Physical and Computational Sciences Directorate} \\
\textit{Pacific Northwest National Laboratory}\\
Richland, Washington, 99352, United States \\
niri.govind@pnnl.gov}
\and
\IEEEauthorblockN{Yuri Alexeev}
\IEEEauthorblockA{ \centerline{ \textit{Computational Science Division}} \\
\textit{Argonne National Laboratory}\\
Lemont, Illinois 60439, United States \\
yuri@anl.gov}
\thanks{$^\dagger$ Equal contributions}}
\maketitle

\begin{abstract}
QuYBE is an open-source algebraic compiler for the compression of quantum circuits. It has been applied for the efficient simulation of the Heisenberg Hamiltonian on quantum computers. Currently, it can simulate the time dynamics of one-dimensional chains. It includes modules to generate the quantum circuits for the above as well as produce the compressed circuits, which are independent of the time step. It utilizes the Yang-Baxter equation (YBE) to perform the compression. QuYBE enables users to seamlessly design, execute, and analyze the time dynamics of the Heisenberg Hamiltonian on quantum computers. QuYBE is the first step toward making the YBE technique available to a broader community of scientists from multiple domains. The QuYBE compiler is available at \href{https://github.com/ZichangHe/QuYBE}{https://github.com/ZichangHe/QuYBE}.
\end{abstract}

\section{Introduction}

Current major efforts towards demonstrating a real quantum advantage are focused on designing practical large-scale quantum computer architectures. To accommodate these efforts with the supporting software, robust quantum compilation processes are much needed. A typical quantum compilation includes three-level circuit generation or optimization, namely, the circuit generation from a reduced set of universal quantum gates, the fault-tolerant design of the original circuit, and the circuit compilation to hardware-specific instructions. There is extensive literature devoted to employing classical approaches for quantum compiling (see Refs. \citenum{Oddi2018Greedy,H_ner_2018,Heyfron_2018,Cincio_2018,Venturelli_2018,Chong2017Programming,Nam2018Automated,Booth2012Quantum,Fowler2011Constructing,Maslov08Quantum,Booth2018Comparing} for past examples using temporal planning, machine learning, and other techniques). Recent exciting ideas include utilizing algebraic relations to compress and optimize the original circuit for more fault-resilient performance on NISQ devices\cite{bassman2021constantdepth,K_kc__2022,camps2021,lin2021real,cirstoiu2020variational,atia2017fast,Berthusen2021,barratt2021parallel}. The Yang-Baxter equation (YBE) or star-triangle relation has recently attracted much attention in quantum information science. This equation, which was originally introduced in theoretical physics\cite{yang1967some} and statistical mechanics\cite{baxter2016exactly}, is a consistency condition which arises in systems where the dynamics are two-body factorizable. The YBE link to quantum computing arises from the connection to topological entanglement, quantum entanglement, and quantum computational universality\cite{ge2016YBE,nayak2008non,kauffman2010topological,zhang2013integrable,vind2016experimental,batchelor2016yang,YBE_pra}. 

Briefly, the quantum YBE can be written as 
\begin{equation}
    (\mathcal{R}\otimes\mathds{1})(\mathds{1}\otimes \mathcal{R})(\mathcal{R}\otimes\mathds{1}) = (\mathds{1}\otimes \mathcal{R})(\mathcal{R}\otimes\mathds{1})(\mathds{1}\otimes \mathcal{R})
\end{equation}
where the $\mathcal{R}$ operator is a linear mapping $\mathcal{R}: V\otimes V \rightarrow V\otimes V$ of vector space $V$. In the present discussion, $\mathcal{R}$ represents a parameterized unitary gate and is limited as a two-qubit gate parameterized by a phase factor and a rotation. Since the YBE connects topological concepts like knots and links to entangled quantum states, the CNOT gate can be replaced by some unitary gate $R$ via the YBE to maintain the universality of quantum computation.\cite{baxter2016exactly} Here, the unitary $R$ gate serves as the solution for the condition of topological braiding, as well as the unitary solution to the YBE.

In our previous work\cite{YBE_pra}, we have proved that for some model systems it is feasible to compress their time evolution circuit based on the YBE to a depth that scales linearly with respect to the number of qubits $N$. Furthermore, the YBE compilation in the classical pre-processing step scales as $\mathcal{O}(N^3)$. Following our previous conceptual work, and with an eye toward large-scale applications, here we report the design and execution of a quantum compiler (QuYBE) based on our previous conceptual YBE analysis. We demonstrate that our QuYBE compiler can successfully perform efficient YBE circuit compression for certain classes of the Heisenberg model.

\section{Methodology}
\begin{figure}[ht]
    \centering
    \includegraphics[scale=0.3]{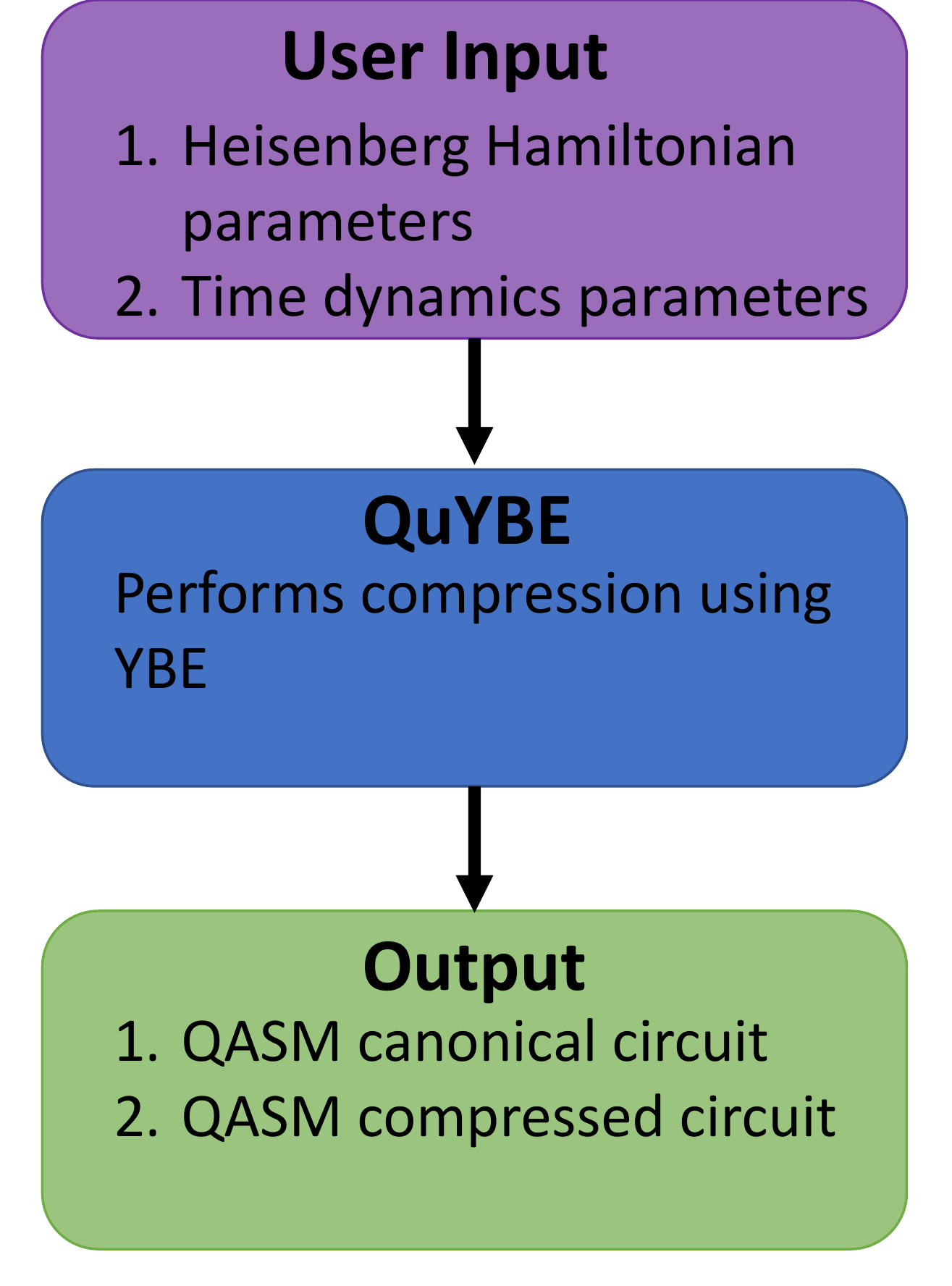}
    \caption{Worflow of QuYBE compiler.}
    \label{fig:workflow}
\end{figure}

% Explain circuit compression in details
% \begin{enumerate}
%     \item Algorithm step by step (1 for core function, 1 for wrapper function)
%     \item Add figures respectively
%     \item Example and how applied to Heisenberg. 
% \end{enumerate}
In this section, as an illustration, we introduce how our QuYBE compiler compresses the quantum circuit corresponding to the time dynamics of the 1D-Heisenberg XY model Hamiltonian. The same method can also be applied to other Heisenberg model Hamiltonians with or without external field. For an $N$-spin 1D-Heisenberg XY model Hamiltonian, to simulate its time dynamics with $T$ Trotterization steps, the corresponding circuit has $M=2T$ layers. For such a circuit with $M>N$, the QuYBE compiler is able to compress the full circuit to $N$ layers analytically without loss of accuracy.
The workflow is summarized in Fig.~\ref{fig:workflow}. 
In the following, we elaborate our algorithms that have been used in the QuYBE compiler. We will first be focused on a sequential compression scheme for a demonstration purpose, and then we will briefly discuss how it can be made parallel.

\subsection{Sequential compression scheme}

\begin{figure}
    \centering
    \includegraphics[width=\linewidth]{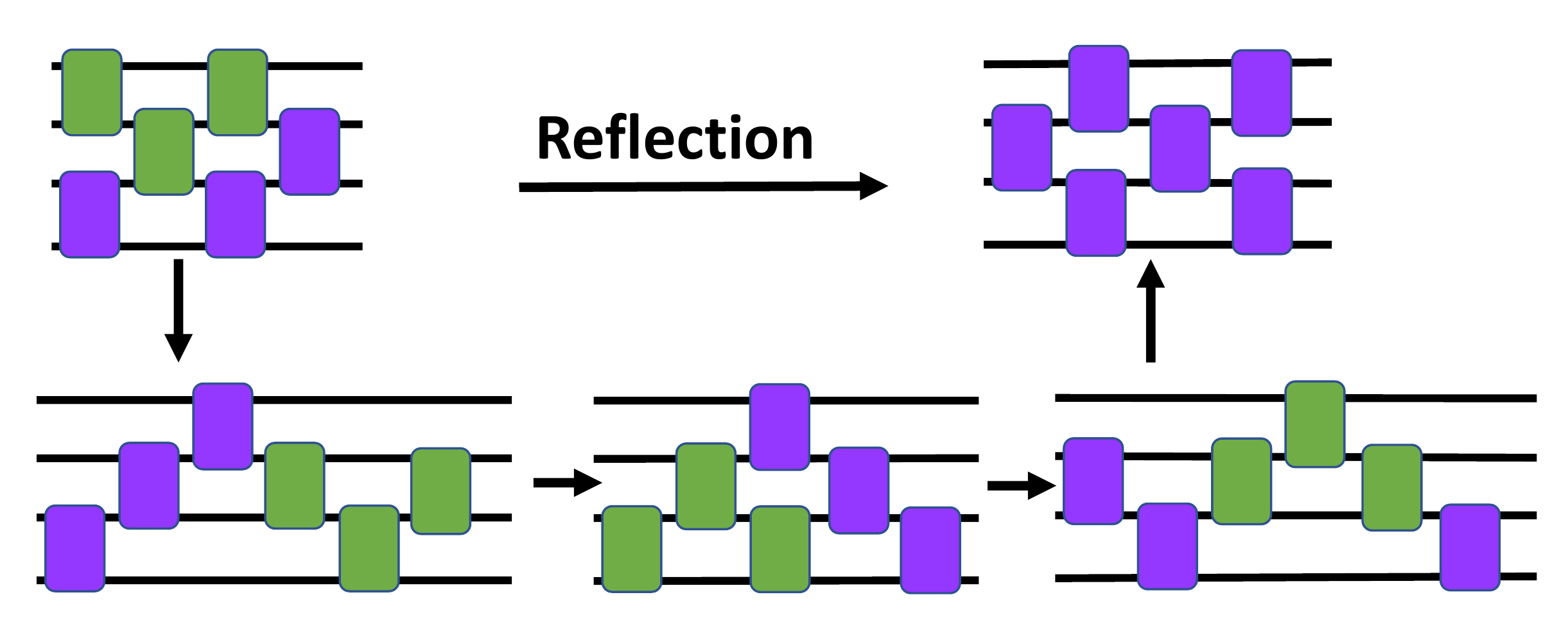}
    \caption{Reflection symmetry can be achieved by using the YBE four times on four qubits. For each step, the triplets of qubits on which YBE act is shown by green color.}
    \label{fig:example_reflection}
\end{figure}

The overall algorithm is described in Algorithm.~\ref{alg:compression}. It first transforms the first $N$-layer circuit to the reflection symmetric form. Then the $(N+1)$-th layer can be merged into the reflected circuit, and the circuit depth is decreased by one. By repeating these two steps, we can compress the original circuit to a compact form with only $N$ layers, i.e. circuit depth is a linear function of number of qubits $N$. As an example how YBE fascilitate the reflection symmetry, Fig \ref{fig:example_reflection} 
shows how to achieve reflection symmetry for four qubits where YBE is true. 
\begin{algorithm}
\caption{A sequential circuit compression employing QuYBE}\label{alg:compression}
\begin{algorithmic}[1]
\Require A $N$-qubit time evolution circuit with $M (M > N)$ layers
\Ensure Functionally equivalent circuit with $N$ layers
\While{$M > N$}
\State Execute the function \textsc{CircuitReflection} on the first $N$ layers. %into the reflection symmetry form based on Algorithm.~\ref{alg:reflection_sym}
\State Merge the adjacent gates on the same qubits %based on the merge function
\State $M \gets M - 1$
\EndWhile
\end{algorithmic}
\end{algorithm}

\begin{figure}[t]
    \centering
    \includegraphics[width=\linewidth]{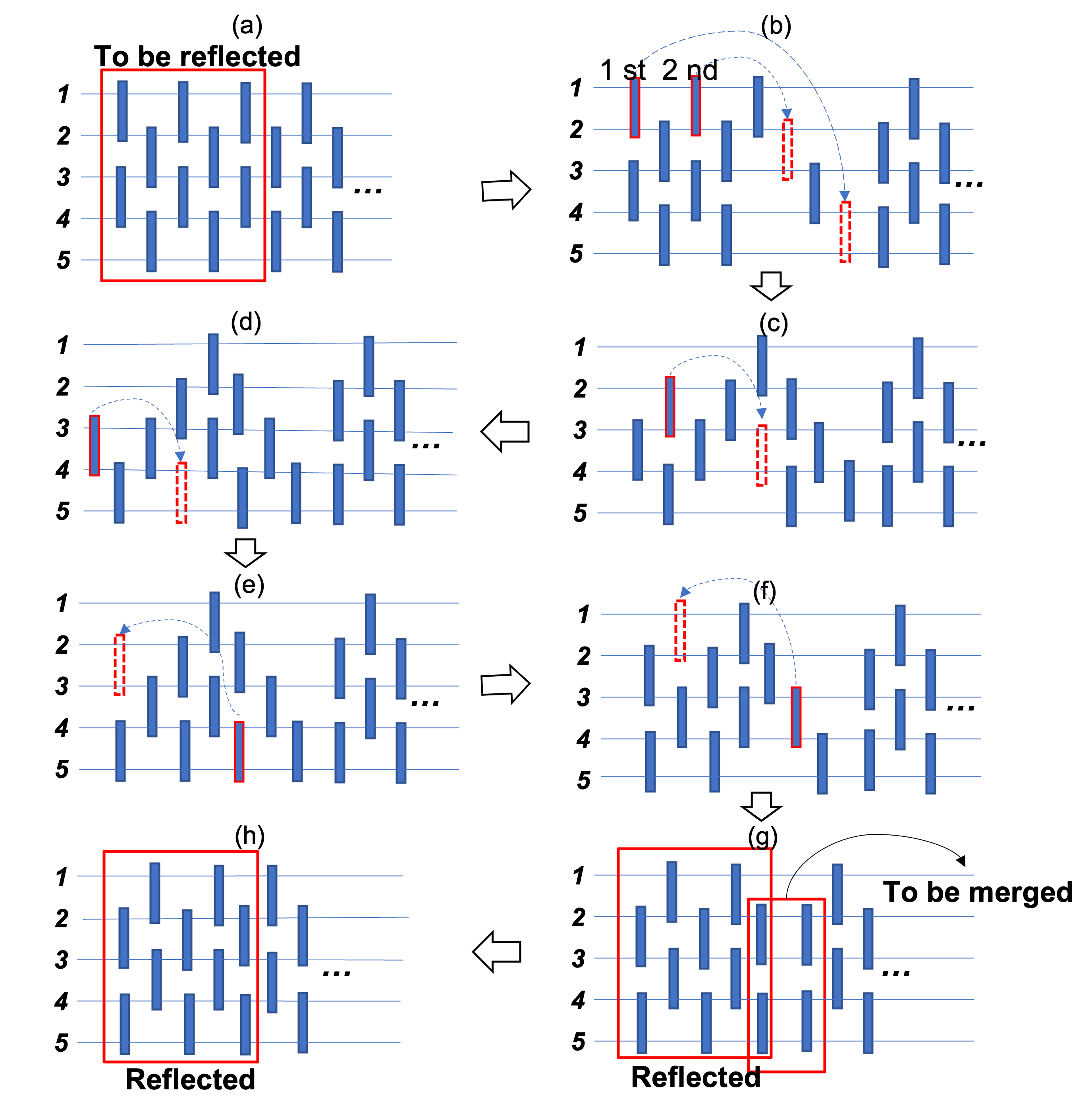}
    \caption{Schematic demonstration of compressing one layer from the five-qubit time evolution circuit. (a) shows the original circuit. (b-g) converts the first five-layer circuit into its reflection symmetric form, where (c-e) reflects the four-layer four-qubit time evolution circuit to its reflection symmetric form. (g) shows the circuit that enables the merge of the fifth and six-th layers. (f) shows the compressed circuit.}
    \label{fig:five_qubit_example}
\end{figure}

Algorithm \ref{alg:compression} simply describes how a sequential compression works for a given time evolution circuit of the Heisenberg XY model. Inside the compression algorithm, the key function $\textsc{CircuitReflection}$ is the multi-qubit reflection operation, of which the detailed workflow is given in Algorithm \ref{alg:reflection_sym}. As can be seen, $\textsc{CircuitReflection}$ is a recursive function, which also reflects how we proved the existence of a general reflection symmetry of an $N$-layer $N$-qubit time evolution circuit for the studied models in our previous analysis. \cite{YBE_pra} Essentially, what $\textsc{CircuitReflection}$ does is to gradually offload the two-qubit gates in a given time evolution circuit from top left to lower right and restore the \textit{untouched} two-qubit gate from lower right to top left to get the reflection done. To better illustrate how exactly the key function $\textsc{CircuitReflection}$ works, take a five-qubit time evolution circuit as an example, Fig. \ref{fig:five_qubit_example} shows how we can offload the restore the two-qubit gates to vertically reflect the first five layers of the circuite to enable the merge between the fifth and sixth layers.

\begin{algorithm}
\caption{Multi-qubit circuit reflection operation}\label{alg:reflection_sym}
\begin{algorithmic}[1]
\Function{\textsc{CircuitReflection}}{circuit of $N$-qubit}
\State Offload the gate on the first two qubits to the side. 
\State Perform \textsc{CircuitReflection} for the circuit of ($N-1$)-qubit.
\State Restore the \textit{untouched} two-qubit gates in Step 1 back to the top row.
\EndFunction
\end{algorithmic}
\end{algorithm}
 
%\begin{algorithm}
%\caption{\textsc{CircuitReflection}}\label{alg:reflection_sym}
%\begin{algorithmic}[1]
%\Require A $N$-qubit time evolution circuit with $N$ layers
%\Ensure Functionally equivalent circuit in the reflection symmetric form
%\State \text{{//}  From time dynamics to general A-shape}
%\For{\texttt{$i = 1$, $i{+}{+}$, and $i < N-1$}}
%    \State Focus on the first $N-i$ square-looked layers
%    \State Move $i$-th row's XY gates to the right and tucked between the two-qubit gates to the last layer based on A2V and V2A functions. 
%\EndFor
%\State \text{{//} From general A-shape to time dynamics}
%\If{$N$ is even}
%\State \texttt{Initial} = 2
%\ElsIf{$N$ is odd}
%\State \texttt{Initial} = 3
%\EndIf
%\For{\texttt{$i$=\text{Initial}, $i{+}{+}$, while $i < N-1$}}
%    \State Focus on the first $i$-th diagonal-looked layers
%    \State Move $i$-th diagonal layer's XY gates back to the top row, i.e. (i-1)-th row of the overall circuit, based on A2V and V2A functions.
%\EndFor
%\end{algorithmic}
%\end{algorithm}

\begin{figure}[ht]
    \centering
    \includegraphics[scale=0.7]{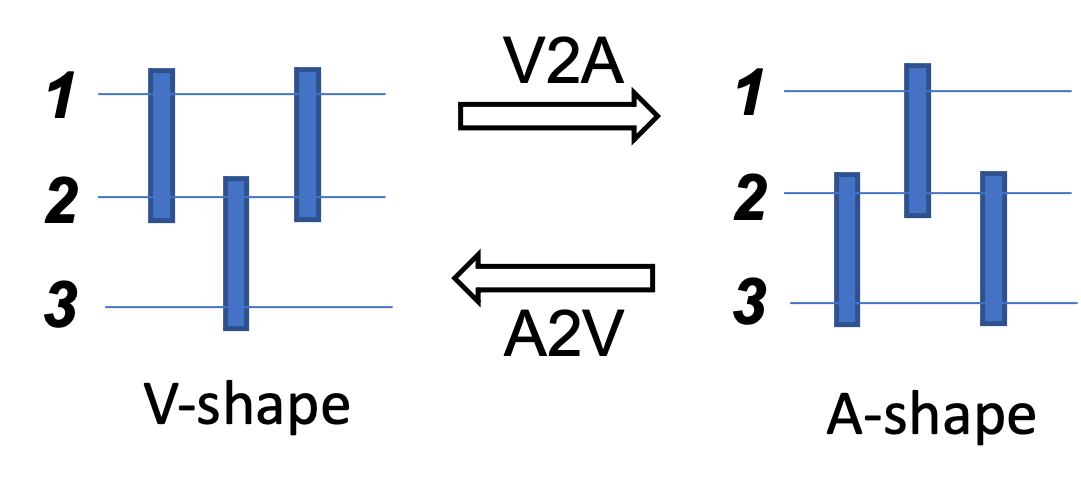}
    \caption{Three-qubit reflection operation is treated as a basic operation in our QuYBE compiler.}
    \label{fig:3qubit_YBE_example}
\end{figure}

Another key component in Algorithm \ref{alg:reflection_sym}, as can be seen from Fig. \ref{fig:five_qubit_example}, is the three-qubit YBE (or reflection) operation which we have taken in our QuYBE compiler as a basic operation. As demonstrated in Fig. \ref{fig:3qubit_YBE_example}, the three-qubit YBE operation includes transforming an A-shape circuit to a V-shape circuit (denoted as the A2V transformation), and its reverse operation (denoted as the V2A transformation).
%we need three core functions (defined in Algorithm.~\ref{alg:YBE_core}): A2V transformation, V2A transformation, and merge. 
%The 3-qubit A- and V-shaped circuits are shown in Fig.~\ref{fig:3qubit_YBE_example}. 
We have coded these two transformations, as well as a gate merge function, as core functions in our QuYBE compiler. 
%The A2V and V2A functions can be seen as the demonstration of the reflection symmetry of a 3-qubit system. 
The detailed analytic transformation rules are based on the Theorem I in Ref.\citenum{YBE_pra}. Also, regarding the scaling of $\textsc{CircuitReflection}$, the reflection of an $N$-qubit time evolution circuit with $N$ layers requires $\mathcal{O}(N^3)$ number of 3-qubit YBE transformations (see Fig.~\ref{fig:cnots} (red), also see the derivation in Appendix A of Ref.\citenum{YBE_pra}).
%Two adjacent 2-qubit gates can be merged into one analytically (Eq.~(11) in Ref.\citenum{YBE_pra}). 

\subsection{Parallel compression scheme}

In comparison with some heuristic approaches, the QuYBE compiler offers a deterministic approach for compressing the time evolution circuits for some model systems. Nevertheless, a sequential version would become a very time consuming pre-processing step given only one layer is compress after $\mathcal{O}(N^3)$ three-qubit YBE operations. To reduce the time-to-solution, we also explore the feasibility of having a parallel scheme. 

\subsubsection{Starting point of the circuit compression}

From the above demonstration, one important observation is that the compression in Fig. \ref{fig:five_qubit_example} starts from the very beginning of the circuit structure where there are no gates prior to the circuit fragment that is about to get compressed. Therefore, we only achieve an ($N+1$)-to-$N$ compression. Alternatively, if we choose to start the compression from anywhere other than the beginning, we will get an ($N+2$)-to-$N$ two-layer compression, i.e., one layer gets compressed at the beginning of the circuit fragment and the other gets compressed at the end of the circuit fragment, after performing one reflection operation (see the inset in Fig. \ref{fig:parallel}). 

\subsubsection{Parallel compression} 
\begin{figure}[h!]
    \centering
    \includegraphics[width=\linewidth]{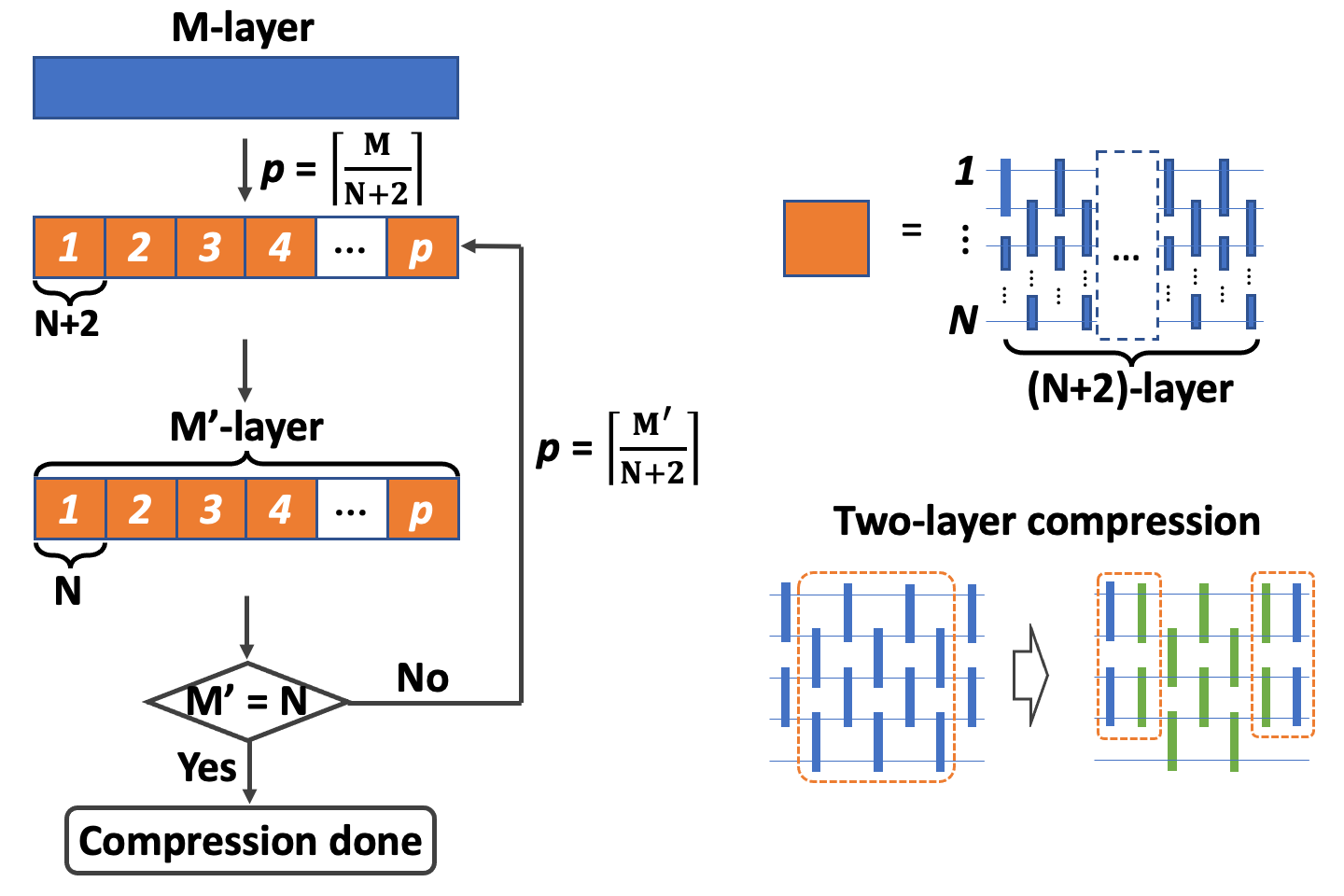}
    \caption{A parallel execution of QuYBE compiler. \textit{p} denotes the maximal number of processes. The inset at the lower right demonstrate a two-layer compression for a seven-layer five-qubit time evolution circuit if the reflection takes place at the layers \#2$-$\#6.}
    \label{fig:parallel}
\end{figure}

Another important observation is that the ($N+2$)-to-$N$ compression in the $(N+2)$-layer is independent of other $(N+2)$-layer circuit fragments. The independence between the circuit fragments then triggers a straightforward parallel scheme to reduce the time-to-solution. Specifically, as shown in Fig. \ref{fig:parallel}, we only need to partition the original circuit to multiple circuit fragments, each of which is an $(N+2)$-layer circuit. The compression of these circuit fragments can straightforwardly be distributed to classical computing resources. Suppose the entire time evolution circuit include $M$ layers, the simple parallel scheme would require $\lceil \frac{M}{N+2} \rceil$ processors at the maximum to do the compression. 

\section{Results}
In this section, we discuss the impact of circuit compression by showing the reduction in two-qubits gate, i.e., CNOTs, and the growth of CNOTs with increasing system size for the time dynamics of 1D-XY Hamiltonian, a subclass of Heisenberg Hamiltonian. 
The Heisenberg Hamiltonian \cite{fazekas1999lecture, skomski2008simple, pires2021theoretical} is widely used to study magnetic systems, where the magnetic spins are treated quantum mechanically. The Hamiltonian (with nearest neighbor interaction) can be written as
\begin{equation}
    \hat{H} = 
    -\sum_{\alpha}\{J_{\alpha}\sum_{i=1}^{N-1} \sigma_{i}^{\alpha}\otimes \sigma_{i+1}^{\alpha}\}, 
\end{equation}\label{eq:Heisenberg}
where $\alpha$ sums over $\{x,y,z\}$, the coupling parameter $J_{\alpha}$ denotes the exchange interaction between nearest-neighbour spins along the $\alpha-$direction, and $\sigma^{\alpha}
_{i}$ is the $\alpha$-Pauli operator on the $i$th
spin. For XY Hamiltonian, $J_z =0$ in Eq. (\ref{eq:Heisenberg}).
We also show a real device (IBM-Manila) simulation using the same compression scheme. 

\begin{figure}[ht]
    \centering
    \includegraphics[width=\linewidth]{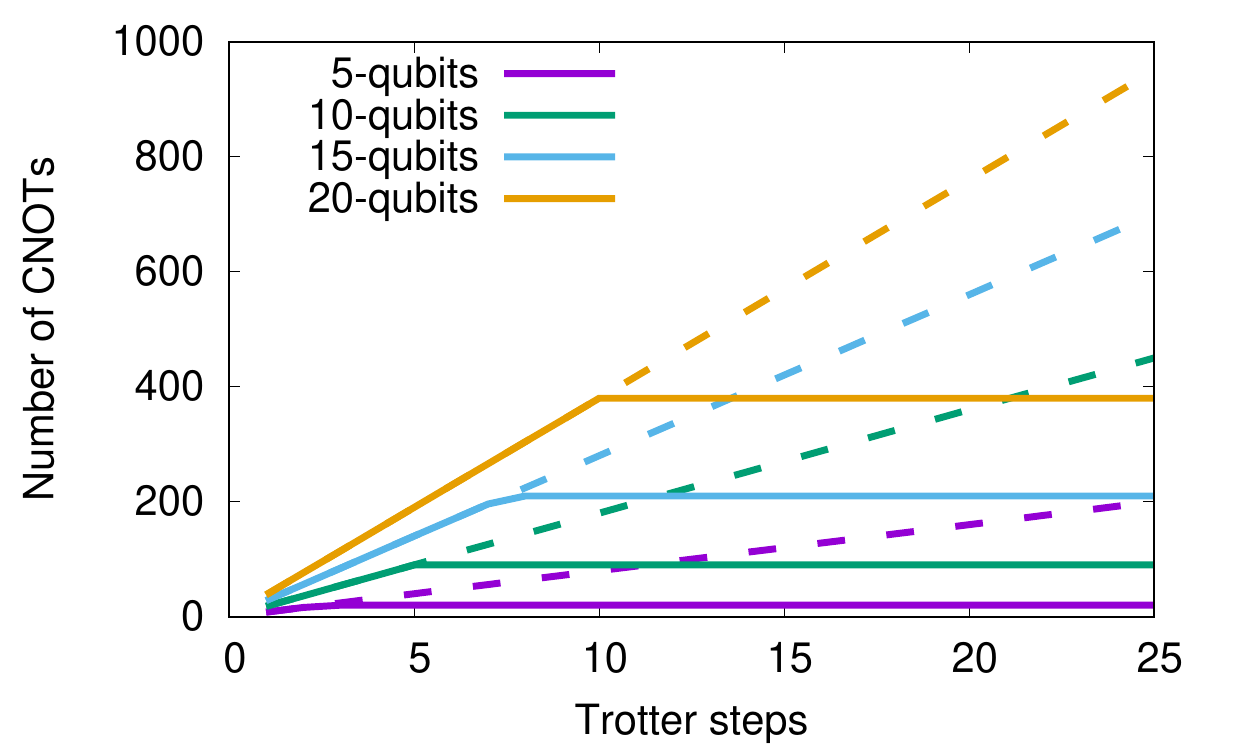}
    \caption{Comparison between the canonical and compressed quantum circuits for the growth of CNOTs with Trotter steps. The dashed line represents the growth of CNOTs in the canonical quantum circuit, while the filled line represents the growth of CNOTs in the compressed quantum circuit. }
    \label{fig:compression}
\end{figure}

\begin{figure}[ht]
    \centering
    \includegraphics[width=\linewidth]{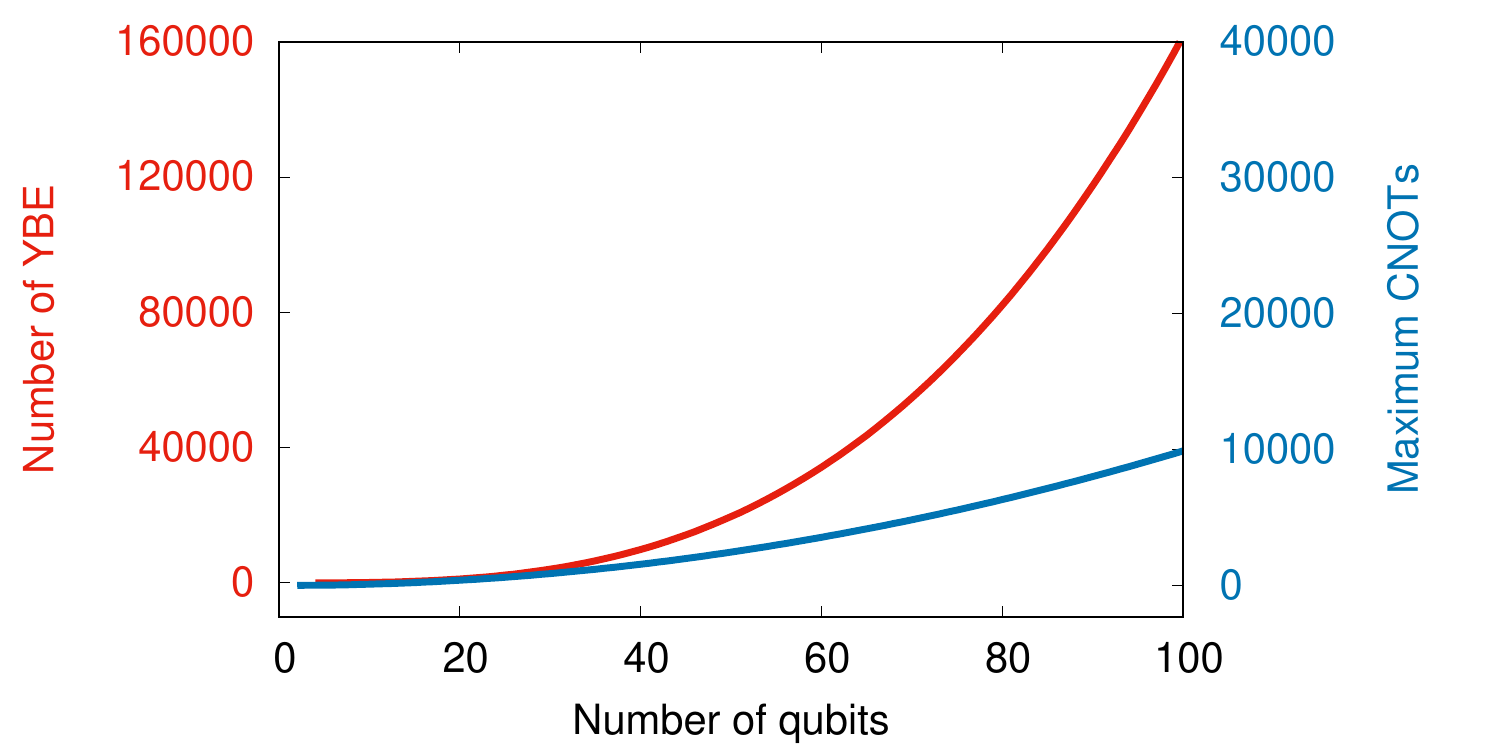}
    \caption{Scaling of number of 3-qubit YBE transformations to finish one reflection operation (red). Scaling of maximum CNOTs is required in compressed quantum circuits for arbitrary time evolution with the number of qubits (blue).}
    \label{fig:cnots}
\end{figure}

\begin{figure}[ht]
    \centering
    \includegraphics[width=\linewidth]{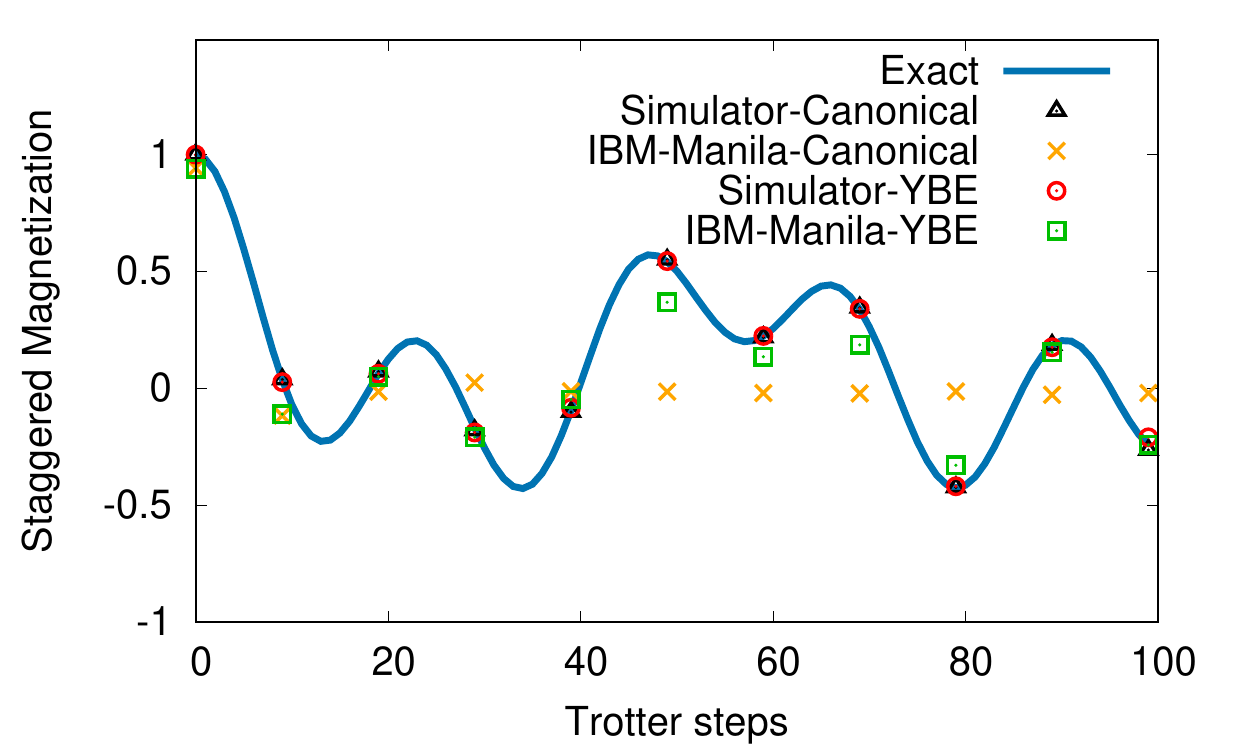}
    \caption{Comparison of the staggered magnetization evolution for 5 spins with the XY Hamiltonian ($J_x = -0.8$, $J_y = -0.2$) between the canonical circuit and compressed circuit via YBE. Each simulation on IBM-Manila used 20000 shots.}
    \label{fig:simulation}
\end{figure}

Fig. \ref{fig:compression} shows the growth of CNOTs between the canonical and compressed circuits with increasing steps in the quantum time dynamics circuit. 
The dashed line represents the growth of CNOTs with Trotter steps in the canonical circuit. On the other hand, after the compression, CNOTs grow until a step number and plateau to a constant after that, as shown by the filled line. Fig \ref{fig:cnots} (blue) shows the scaling of the maximum CNOTs required for arbitrary time dynamics in compressed circuits with the number of qubits. 
The growth of CNOTs is quadratic with the number of qubits. Fig \ref{fig:simulation}, shows the evolution of the staggered magnetization for 5-spins for the Heisenberg XY Hamiltonian ($J_x = -0.8$, $J_y = -0.2$). Each Trotter step is 0.05 units of time. 
%The very evolution act as a baseline. 
In order to confirm the validity of compression, we plot results from the canonical circuit and compressed circuit using the Qiskit simulator for every tenth step. It shows that the compressed circuit (composed of only 20 CNOTs) provides the same level of accuracy as the canonical circuit on the simulator. 

The impact of the compression is clearly seen when the same simulation is run on a real quantum device. 
Fig. \ref{fig:simulation} shows the results of the canonical circuit and compressed circuit on the IBM-Manila quantum device. We computed the time-dependent staggered magnetization, $m_s(t)$, which can be connected to the antiferromagnetism and ferrimagnetism in materials as follows:
\begin{equation}
    m_s(t) = \frac{1}{N} \sum_{i} (-1)^{i}\braket{\sigma_{z}(t)}. 
\end{equation}
%of the system with time, which is directly connected to %anti-ferromagnetic parameter of materials. 
The initial state is the ground state (Ne\'el state) of the XY Hamiltonian, defined as $\Psi_{0} = \ket{\uparrow\downarrow\uparrow\downarrow...\uparrow\downarrow}$. The staggered magnetization of the Ne\'el state is one. We performed the time evolution for 5 units of time with a Trotter step size of 0.05 units.

The staggered magnetization using the canonical circuit quickly drops to zero and remains zero for future time steps. The increasing circuit depth and noise results in a maximally mixed state corresponding to zero magnetization. This does not happen in the compressed circuit, which only requires 20 CNOTs for arbitrary time evolution. In other words, the maximum error is capped, and we observe better qualitative and quantitative results on real quantum devices.

\section{Conclusions}
In this paper, we have presented the QuYBE, which is an open-source algebraic compiler for the compression of quantum circuits. It has been applied for the efficient simulation of the time dynamics of the Heisenberg Hamiltonian on quantum computers. The compiler is based on a circuit compression algorithm based on the quantum Yang-Baxter equation. We demonstrate the efficacy of this approach by comparing time dynamics simulations with canonical and compressed circuits of the 1D-Heisenberg XY model on a real quantum device. 

The QuYBE compiler is available at \href{https://github.com/ZichangHe/QuYBE} {https://github.com/ZichangHe/QuYBE}. One of our goals is to make the QuYBE compiler easy to use with existing popular quantum compilers. It is achieved by using QuYBE as an intermediate step in the compilation pipeline, where YBE takes a QASM circuit, compresses it, and generates a QASM circuit in the target format.

The proposed YBE technique is a general approach that can be applied to any quantum circuit compliant partially or fully with YBE. For example, other types of Ising Hamiltonians are obvious targets. This work is important because it paves the way for large-scale applications and investigations of broader classes of models.

\section*{Acknowledgment}
This material is based upon work supported by the U.S. Department of Energy, Office of Science, and National Quantum Information Science Research Centers. Y.A. acknowledges support from the U.S. Department of Energy, Office of Science, under contract DE-AC02-06CH11357 at Argonne National Laboratory.

\bibliographystyle{IEEEtran}
\bibliography{reference}

\end{document}